\documentclass[
 amsmath,amssymb,
 aps,
 pra,twocolumn
]{revtex4}

\usepackage{verbatim,color}
\usepackage{graphicx}
\usepackage{dcolumn}
\usepackage{bm}

\newcommand{\ket}[1]{\ensuremath{\left|#1\right\rangle}}
\newcommand{\bra}[1]{\ensuremath{\left\langle#1\right|}}

\newcommand{\ketbra}[2]{\ket{#1}\!\!\bra{#2}}
\newcommand{\braopket}[3]{\ensuremath{\bra{#1}#2\ket{#3}}}
\newcommand{\proj}[1]{\ketbra{#1}{#1}}
\def\Id{1\!\mathrm{l}}
\newcommand{\Tr}{\mathrm{Tr}}
\newcommand{\expec}[1]{\ensuremath{\left\langle#1\right\rangle}}

\newcommand{\rrangle}{\rangle\!\rangle} \newcommand{\llangle}{\langle\!\langle}
\newcommand{\sket}[1]{\ensuremath{\left|{\scriptstyle #1}\right\rrangle}}
\newcommand{\sbra}[1]{\ensuremath{\left\llangle{\scriptstyle #1}\right|}}
\newcommand{\sbraket}[2]{\ensuremath{\left\llangle{\scriptstyle #1}|{\scriptstyle #2}\right\rrangle}}
\newcommand{\sketbra}[2]{\sket{#1}\!\!\sbra{#2}}
\newcommand{\sbraopket}[3]{\ensuremath{\sbra{#1}#2\sket{#3}}}
\newcommand{\sproj}[1]{\sketbra{#1}{#1}}

\begin{document}

\title{On the Optimal Choice of Spin-Squeezed States for Detecting and Characterizing a Quantum Process}
 
\author{Lee A. Rozema$^1$}
 
\author{Dylan H. Mahler$^1$}

\author{Robin Blume-Kohout$^2$}
 
\author{Aephraim M. Steinberg$^1$}

\affiliation{%
$^1$Centre for Quantum Information \& Quantum Control and Institute for Optical Sciences,
Dept. of Physics, 60 St. George St., University of Toronto, Toronto, Ontario, Canada M5S 1A7\\
$^2$Computer Science Research Institute (CSRI), Sandia National Laboratories, Albuquerque NM 87123
}

\date{\today}
\begin{abstract}
Quantum metrology uses quantum states with no classical counterpart to measure a physical quantity with extraordinary sensitivity or precision.  Most metrology schemes measure a single parameter of a dynamical process by probing it with a specially designed quantum state.  The success of such a scheme usually relies on the process belonging to a particular one-parameter family.  If this assumption is violated, or if the goal is to measure more than one parameter, a different quantum state may perform better.  In the most extreme case, we know nothing about the process and wish to learn everything.  This requires quantum process tomography, which demands an informationally-complete \emph{set} of probe states.  It is very convenient if this set is \emph{group-covariant} -- i.e., each element is generated by applying an element of the quantum system's natural symmetry group to a single fixed fiducial state.  In this paper, we consider metrology with 2-photon (``biphoton'') states, and report experimental studies of different states' sensitivity to small, unknown collective $SU(2)$ rotations (``$SU(2)$ jitter'').  Maximally entangled N00N states are the most sensitive detectors of such a rotation, yet they are also among the \emph{worst} at fully characterizing an a-priori unknown process.  We identify (and confirm experimentally) the best $SU(2)$-covariant set for process tomography; these states are all \emph{less} entangled than the N00N state, and are characterized by the
fact that they form a 2-design.
\end{abstract}

\maketitle

\section{Introduction}
The goal of quantum metrology is to measure or detect physical phenomena with surprising precision by exploiting quantum resources.  Often, this means using entangled states to achieve greater resolution or sensitivity.  For example, squeezed light\cite{breitenbach_measurement_1997} and N00N states \cite{boto_quantum_2000, mitchell_super-resolving_2004,walther_broglie_2004} have been used in interferometers to achieve higher precision in \emph{single parameter} estimation.  N00N states are maximally sensitive to small $U(1)$ phase shifts \cite{giovannetti_advances_2011}, but they are fragile.  Other parameters might be best detected or estimated by a different optimal state \cite{crowley_multiparameter_2012,humphreys_quantum_2013}, and for estimating a even a simple three-parameter SU(2) process the optimal state is unknown \cite{zhou_quantum-enhanced_2014}.
At the opposite extreme of metrology is \emph{quantum process tomography} (QPT) \cite{mitchell_diagnosis_2003, obrien_quantum_2004}.  Here, the goal is to learn \emph{every} parameter of an unknown process.  QPT requires a diverse set of probe states, and the overall accuracy of estimation depends on the properties of the entire set.  For process tomography on a single quantum optical mode, Lobino {\it et al.} \cite{lobino_complete_2008} showed that it is sufficient to (1) prepare a single Glauber coherent state and (2) displace it by a variety of phase space translations.  This approach, in which a single ``fiducial'' state is multiplied into a complete set of probe states by easily implemented group transformations, has the great merit of experimental ease.  But while sufficiently large coherent-state ensembles are sufficient for process tomography, they are not efficient.  Coherent states are very ``classical'' \cite{glauber_coherent_1963}, and provide exponentially little information about parameters of some quantum processes,
motivating a search for set of states for tomography which provide equal information about all possible processes (see \cite{blume-kohout_curious_2011} for a precise statement of this problem).

In this paper, we examine a closely related question for 2-photon polarization (``biphoton'') states.  Like an optical mode, this system admits [spin]-coherent states (as well as others).  The corresponding symmetry group, $SU(2)$, is transitive on the set of coherent states, i.e. a spin-coherent state can be transformed into any other spin-coherent state by applying a polarization (SU(2)) rotation.  We prepare a wide range of probe states, and quantify their performance at two opposite extremes of the metrology spectrum: (1) their ability to \emph{detect} random $SU(2)$ phase shifts, and (2) their ability to \emph{characterize} an unknown process, when displaced by a variety of $SU(2)$ rotations and used for QPT.  Remarkably, the most sensitive detector states (N00N states) are also among the \emph{least} effective for QPT!  The optimal $SU(2)$-covariant set (i.e. a set generated by applying SU(2) operations to a fiducial state) for QPT is generated by a state that is neither spin-coherent nor N00N, but outperforms both of them.  When displaced by uniformly random $SU(2)$ operations, it generates a 2-design \cite{scott_optimizing_2008, fernandez-perez_quantum_2011}, confirming the theoretical prediction that 2-designs should be optimal for process tomography.

We focus on a particular, important family of processes that we call \emph{$SU(2)$ jitter}.  In $SU(2)$ jitter, an $N$-photon state experiences a small random collective $SU(2)$ rotation, whose magnitude is Gaussian-distributed.  Detecting and characterizing $SU(2)$ jitter is important because it is a common model for \emph{decoherence}\cite{Rivas_SU2_2013}, the primary enemy of quantum information and computation \cite{mahler_identification_2012}.  Noiseless subsystems \cite{viola_dynamical_2000, viola_experimental_2001} were designed against this noise model.  

\begin{figure*}
\includegraphics[scale=.6]{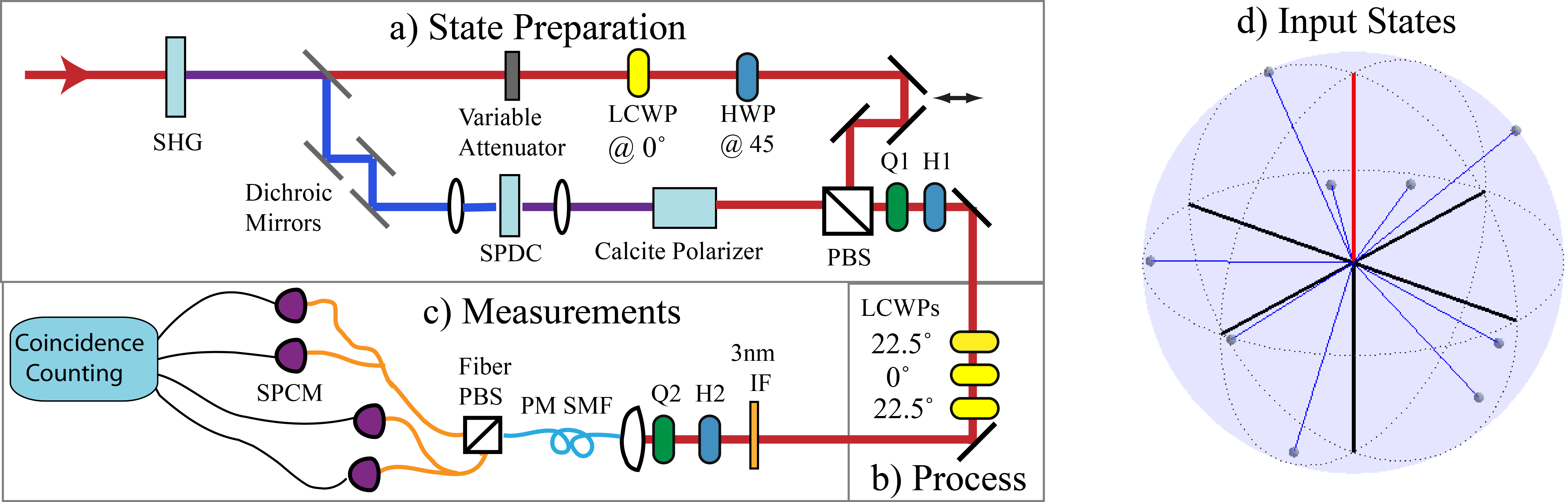}
\caption{\label{fig:1} Schematic diagram of the experimental apparatus used to generate and measure different biphoton states. {\bf a) State Preparation} -- We can prepare any biphoton state in two steps.  First, we prepare a state of the form $\sqrt{x}\ket{2,0}_{H,V}+e^{i\phi}\sqrt{1-x}\ket{0,2}_{H,V}$ by using a polarizing beam splitter (PBS) to combine a vertically polarized weak coherent state with the output of horizontally polarized type-I collinear down-conversion.  The angle $\phi$ is set by the relative phase between the two paths, which we control using a liquid-crystal wave plate (LCWP).  Then, we apply any desired polarization rotation in $SU(2)$, using quarter- and half-waveplates, to produce any desired biphoton state.  {\bf b) Process} -- 3 LCWPs, oriented as shown, are used to perform arbitrary polarization rotations. To implement decoherence, the retardances of the LCWPs are made to fluctuate during each measurement. {\bf c) State Measurement} -- The biphoton states pass through wave plates, are coupled into a polarization-maintaining fiber and sent to a polarizing beam splitter. The output of each port of the beam splitter is probabilistically split, using 50:50 fiber beam splitters, and sent to single-photon counting modules. {\bf d) Input States} -- A graphical representation of the set of states used for process tomography.  A fiducial state is prepared and rotated to nine other states.  These states (including the unrotated fiducial state) make up the set of ten input states used for process tomography.  The fiducial state is represented by the red line, and it is rotated to each of the nine other points on the sphere.  The rotations are chosen to be (approximately) uniformly distributed on the surface of the sphere.}
\end{figure*}

\section{Experimental Methods and Preliminary Validations}

A \emph{biphoton} is a system of two photons in the same spatial and temporal modes\cite{bogdanov_qutrit_2004, adamson_multiparticle_2007}, and its polarization is isomorphic to a spin-1 particle (since each photon polarization is isomorphic to a spin-1/2 particle) \cite{adamson_detecting_2008}.  Basis states can be described by specifying the number of photons polarized horizontally (H) and vertically (V), e.g.
\begin{equation}
C_0\ket{2,0}_{H,V} + C_1\ket{1,1}_{H,V} + C_2\ket{0,2}_{H,V},
\end{equation}
or, in the basis of spin states $\ket{J,m}$,
\begin{equation}
C_0\ket{1,+1} +C_1\ket{1,0} +C_2\ket{1,-1}.
\end{equation}
A collective polarization rotation corresponds to an $SU(2)$ rotation of the effective spin-1 particle about some axis $\vec{r}$ by an angle $\theta$.  Two photons polarized in the same direction form a \emph{spin-coherent} state, and these spin-coherent states are analogous to the Glauber coherent states of an optical mode \cite{arecchi_atomic_1972}.  Just as \cite{lobino_complete_2008} used displaced Glauber coherent states as input states for QPT, a set of at least nine distinct spin-coherent states can form a complete probe set for QPT, and can be generated by applying various $SU(2)$ rotations to a single fiducial spin-coherent state.  Here, we generalize this procedure in a simple way:  we prepare a fiducial state that is \emph{not} spin-coherent, and generate candidate probe sets for QPT by applying 10 distinct $SU(2)$ rotations to it.  (We prepare 10 fiducial states instead of 9 because it is experimentally convenient, and provides a small amount of useful redundancy).  Our fiducial states take the form 
\begin{equation}\label{eq:psi}
\ket{\psi_x} = \sqrt{x}\ket{2,0}_{H,V}+\sqrt{1-x}\ket{0,2}_{H,V},
\end{equation}
and are prepared using the apparatus sketched in Fig. \ref{fig:1}a and the methods described in \cite{hofmann_high-photon-number_2007,fiurasek_conditional_2002, afek_high-noon_2010,rosen_sub-rayleigh_2012,israel_experimental_2012, rozema_scalable_2013}.  This class of states includes spin-coherent states ($x=0,1$) and the two-photon N00N state ($x=\frac12$).  In fact, \emph{any} biphoton state can be prepared by choosing some value of $x$ and then applying some $SU(2)$ rotation using wave plates.  ($N$-photon states have $N-1$ parameters that are $SU(2)$-invariant; for the biphoton, $x$ is the only parameter.)

\begin{figure}
\includegraphics[width=3.5in]{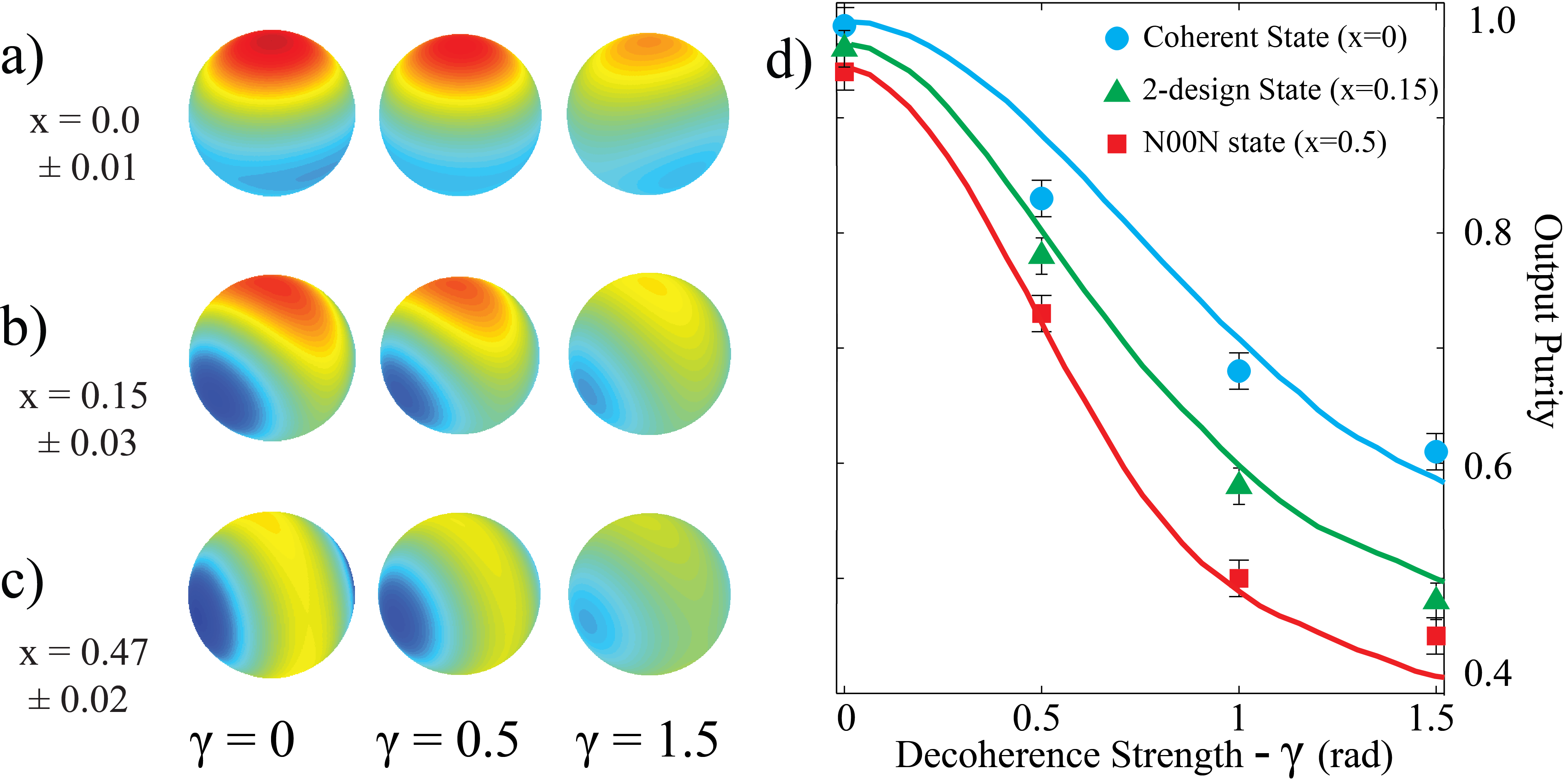}
\caption{\label{fig:2} Typical results of state tomography, and the effect of $SU(2)$-jitter decoherence.  In plots {\bf (a)-(c)}, we show experimentally reconstructed Wigner functions (plotted on the Poincare sphere) for three different states after they have been decohered by three different amounts of $SU(2)$ jitter.  Row (a) shows spin-coherent states with x=0, row (b) shows ``2-design'' states with x=0.15, and row (c) shows nearly-N00N states with x=0.47.  Each row shows the effect of applying decoherence with strength (see Eq. \ref{eq:D}) $\gamma=0,0.5,1.5$.  In {\bf (d)}, we plot the \emph{purity} of the same reconstructed states shown in (a-c).  The solid lines are the theoretical predictions given by simulations of the process (Eq. \ref{eq:D}).}
\end{figure}

The horizontally- and vertically-polarized photons are not perfectly mode-matched when generated.  We remedy this by passing them through a 3nm filter and coupling them into a single-mode fiber, which discards any photons not in the desired mode.  This procedure results in near-perfect biphotons, as quantified by the [very small, $<2\%$] amount of population in the anti-symmetric subspace.  We characterize our state preparation by doing quantum state tomography using the apparatus of Fig \ref{fig:1}c as described in Ref. \cite{adamson_multiparticle_2007, adamson_detecting_2008}.  The resulting experimentally-measured biphoton states, after numerically filtering out the anti-symmetric subspace, are depicted in the first column of Fig. \ref{fig:2} as Wigner distributions plotted on the Poincare sphere \cite{shalm_squeezing_2009}.  As $x$ is increased, the states become more ``nonclassical'', with the most nonclassical state being the N00N state with $x=0.5$ (See row (c), column 1 of Fig. \ref{fig:2}).  This procedure lets us prepare any desired state with fidelity $\geq 93\%$.

The process we study, fluctuating $SU(2)$ rotations, manifests itself in many experimental systems -- e.g., a spin in a fluctuating magnetic field, or a polarization state propagating through a thermally-fluctuating optical fiber.  We consider isotropic decoherence, meaning there is no preferred rotational axis.  The quantum process is:
\begin{equation} \label{eq:D}
D_\gamma[\ket{\psi}\bra{\psi}] = \int{d\vec{r}}\int{d\theta}P(\theta)e^{-i\theta\frac{\vec{r}\cdot\vec{J}}{\hbar}}\ket{\psi}\bra{\psi}  e^{+i\theta\frac{\vec{r}\cdot\vec{J}}{\hbar}},
\end{equation}
where the rotation axis $\vec{r}$ is uniformly random and the angle $\theta$ has a Gaussian distribution
\begin{equation}
P(\theta)\propto~{e^{-\frac{\theta^2}{2\gamma^2}}} 
\end{equation}
The overall strength of the decoherence process is quantified by $\gamma$, the width of the distribution of $\theta$.

We implement the decoherence process described by Eq. \ref{eq:D} using three liquid crystal wave plates (LCWP) as shown in Fig. \ref{fig:1}b.  Each LCWP applies an adjustable polarization rotation, and the rotation angle can be changed very rapidly.  This allows us to apply approximately 50 different (randomly selected) rotations over the timescale of a single process, which closely approximates the ideal process of equation \ref{eq:D} as:
\begin{equation}\label{eq:D2}
\tilde{D}_\gamma[\ket{\psi}\bra{\psi}] = \sum\limits_{k=1}^{50} e^{-i\theta_k\frac{\vec{r}_k\cdot\vec{J}}{\hbar}}\ket{\psi}\bra{\psi}  e^{+i\theta_k\frac{\vec{r}_k\cdot\vec{J}}{\hbar}}.
\end{equation}
Using this implementation we can apply precisely calibrated $SU(2)$ jitter.  We verified both our state preparation \emph{and} our implementation of decoherence by performing state tomography on $\rho = \tilde{D}_\gamma\left[\proj{\psi_x}\right]$ for several values of $x$ and $\gamma$.  Figure \ref{fig:2}a-c shows the Wigner functions of the resulting reconstructed states.  Plots in column 1 are for undecohered states, while columns 2-3 show the effects of $SU(2)$ jitter with strengths of $\gamma=0.5$ and $\gamma=1.5$ rad (respectively).  Increasing $\gamma$ blurs the Wigner function.  This is captured quantitatively by the state's \emph{purity}, a reasonable proxy for the amount of decoherence suffered.  Figure \ref{fig:2}d plots the output purity (computed from the tomographic estimate) versus $\gamma$, and compares it to the prediction of numerical simulations of $\tilde{D}_\gamma$ (solid lines), for three different input states with $x=0$ (blue), $x=0.15$ (green) and $x=0.47$ (red).  The only inputs to our simulation are the experimentally measured purities of the input states when $\gamma=0$, which in a perfect experiment would be 1, but are slightly degraded by experimental noise.  We observe excellent agreement between simulation and experiment, confirming that our process performs as expected.  In particular, the N00N state loses purity more rapidly than any other state (as $\gamma$ is increased), indicating that N00N states are indeed the most fragile (and thus potentially sensitive) to $SU(2)$ jitter.  
The fragility of N00N states to a similar model of SU(2) decoherence was also pointed out in \cite{Rivas_SU2_2013}.  In the next section we discuss an experiment exploiting this fragility to detect decoherence.

\begin{figure}
\includegraphics[scale=.5]{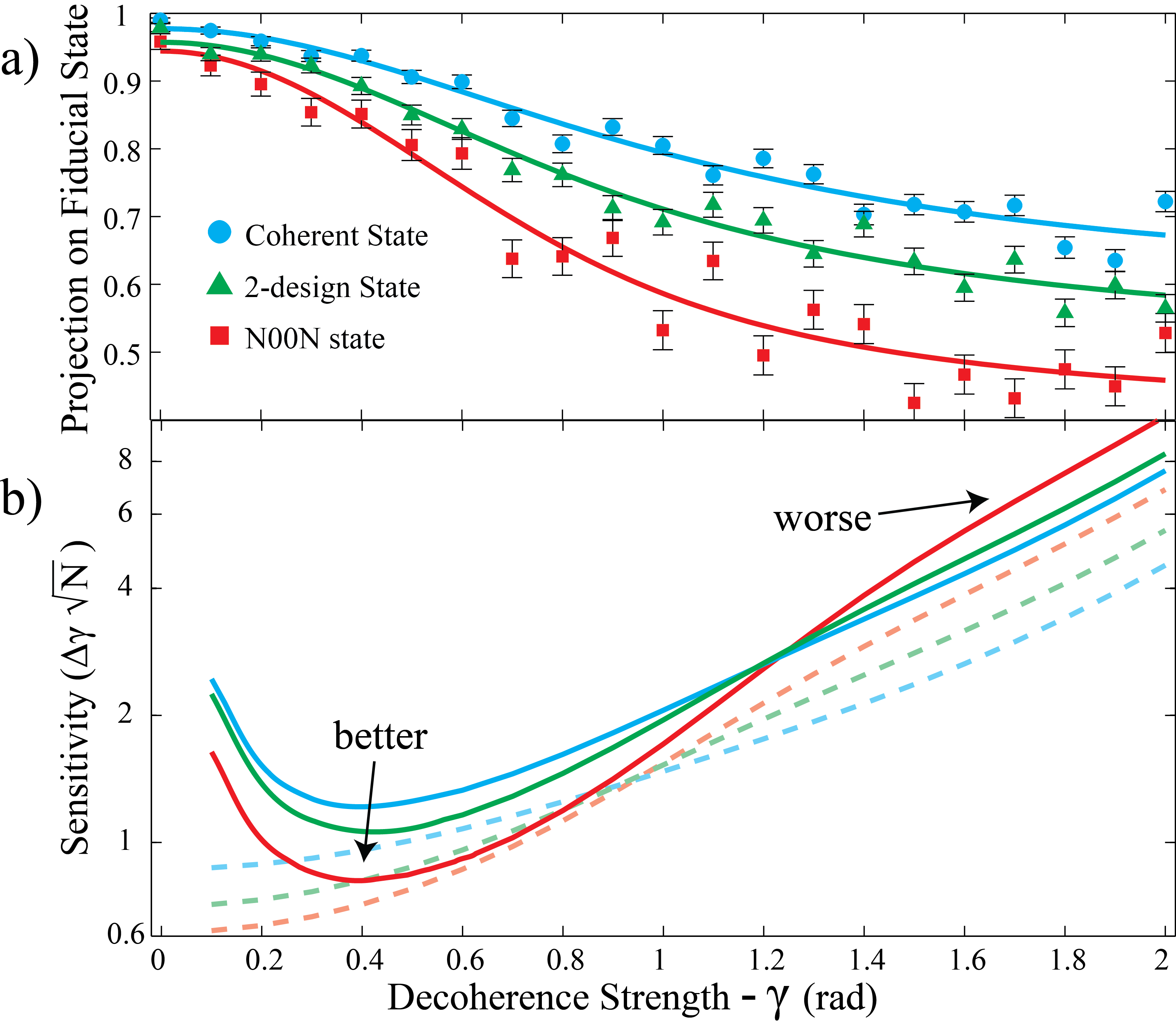}
\caption{\label{fig:3}\textbf{Sensitivity of different states to decoherence:} This figure shows two different measures of the probe state's ability to detect decoherence, for three different probe states, and compares theoretical predictions to experimental data.  Plot {\bf (a)} shows the probability that decoherence is \emph{not} detected, which is simply the probability of finding the system in the same state in which was prepared despite decoherence having happened.  The theory (solid line) is simply the projection $\braopket{\psi_x}{\rho}{\psi_x}$, where $\ket\psi_x$ is the probe state and $\rho$ is the decohered probe state.  Experimental data points are empirical probabilities of nondetection. Plot {\bf (b)} shows the \emph{sensitivity} (Eq. \ref{eq:sensitivity}) of each state to small changes in decoherence, the solid lines are calculated from the slopes of the fit to the detection data shown in (a), and the dashed lines are calculated for ideal input states.}
\end{figure}

\section{Detecting Decoherence}

We examined different probe states effectiveness at \emph{detecting} $SU(2)$ jitter.  This corresponds to distinguishing between two processes:  $\Id$ (no decoherence) or $D_\gamma$ ($SU(2)$ jitter).  Acting on the probe state $\proj{\psi_x}$, these alternatives produce either $\proj{\psi_x}$ or $\rho_D = D_\gamma\left[\proj{\psi_x}\right]$, and to distinguish these alternatives we simply perform a POVM measurement with two outcomes,
\begin{equation} \label{eq:M}
\mathcal{M} = \{\proj{\psi_x},\Id-\proj{\psi_x}\}.
\end{equation}
In simple terms, we are checking to see whether the probe state changed at all.  If done perfectly, this protocol has one-sided error; it may fail to detect $D$, but will never detect it in error.  

To implement this measurement experimentally, we recall that if $\ket{\psi_x}$ is a spin-coherent state then it can be written as $\hat{U}\ket{2,0}_{H,V}$ for some $\hat{U}\in SU(2)$. We can implement $\hat{U}^\dag$ using the quarter- and half-wave plates labeled Q2 and H2 in Fig. \ref{fig:1}c, and after performing this inverse rotation on the output state, detection of two photons at the H-port of the PBS corresponds to the $\proj{\psi_x}$ outcome of $\mathcal{M}$.  Similarly, if $\ket{\psi_x}$ is a N00N state, it can be written as
\begin{eqnarray*}
\ket\psi_x &=& \frac{1}{\sqrt{2}}\left(\ket{2,0}_{H,V}+\ket{0,2}_{H,V}\right) \\
&=& U\ket{1,1}_{H,V}
\end{eqnarray*}
where $U^\dag$ can be implemented by a HWP at $22.5^\circ$.  Thus, after performing this inverse rotation on the output state, detection of 2 coincident photons at the H- and V-ports of the PBS corresponds to the $\proj{\psi_x}$ outcome of $\mathcal{M}$.  

Although it is possible to directly perform this projection for any biphoton state (this could be done, for example, by time reversing the state preparation techniques of \cite{mitchell_super-resolving_2004}), in our experiment, for states with $x\neq 0$ or $x\neq 0.5$, we estimated the value $\expec{\proj{\psi_x}}$ indirectly from two measurements.  In general the density matrix describing the system at the output will be:
\begin{equation}
    \rho = \left(\begin{array}{ccc}
				a            & f & d \\
				f^*            & b & g \\
				d^* & g^* & c 
     \end{array}\right).
\end{equation}
If the process is pure SU(2) decoherence, and the input state is given by equation \ref{eq:psi} then $d$ will be real. (We check that this is the case by performing quantum state tomography on $\rho$ for several decoherence strengths.)  In this case, the expectation value of the projection onto $\ket{\psi_x}$ is
\begin{equation}
\expec{\proj{\psi_x}}=\Tr{(\proj{\psi_x}\rho)} = ax + c(1-x) + 2d\sqrt{x(1-x)}.
\end{equation}
So we can estimate $\expec{\proj{\psi_x}}$, for any value of $x$, by measuring $a$, $c$, and $d$.  If $\rho$ is sent directly to a PBS both photons will be transmitted with probability $a$, both will be reflected with probability $c$, and one will exit each port with probability $b$. All of these probabilities can readily be measured via coincident detection between different combinations of the four detectors in fig \ref{fig:1}c.
To measure $d$, a half-waveplate at $22.5^\circ$ is inserted before the PBS.  Now one photon will exit each port of the PBS with probability
\begin{equation}
P_{HV}=\frac{1}{2}-d-\frac{b}{2}.
\end{equation}
Since $b$ is already known, measuring $P_{HV}$ gives us an estimate of $d$, which gives us enough information to reconstruct $\expec{\proj{\psi_x}}$ for any value of $x$.

We note in passing that this protocol is reminiscent of atomic interferometry.  There, too, a probe state is prepared and then measured later.  The probability (and statistics) of the results typically oscillate over time, because different atomic states have different energies and accumulate quantum phases that beat against one another.  Decoherence makes these oscillations decay and eventually disappear, and this decay is often used to estimate the decoherence strength in the system.  Thus, in both interferometry and our experiment, greater fragility to decoherence enables detection of weaker decoherence.  

We created a range of probe states parameterized by $x\in[0,\frac12]$, varying from a spin-coherent state at $x=0$ to a N00N state at $x=\frac12$ (Fig. \ref{fig:1}a), subjected them to decoherence (Fig. \ref{fig:1}b), and performed the measurement $\mathcal{M}$ given in Eq. \ref{eq:M} (Fig. \ref{fig:1}c).

Figure \ref{fig:3}a shows the experimentally observed probability of failing to detecting $SU(2)$ jitter as a function of the jitter's strength ($\gamma$), for three representative probe states: a N00N state (red), a spin-coherent state (blue) and an intermediate state with $x=0.15$ (green).  The experimental results are in good agreement with the simple theoretical prediction (solid lines) given by
\begin{eqnarray}\label{eq:pGamma}
P(\gamma) &=& \braopket{\psi_x}{D\left[\proj{\psi_x}\right]}{\psi_x} \nonumber \\
 &=& A_x e^{-2\gamma^2} + B_x e^{-\frac{\gamma^2}{2}} + C_x \label{eq:Ptheory},
\end{eqnarray}
where $A_x$, $B_x$ and $C_x$ are straightforward but unwieldy functions of $x$ (see equation \ref{eq:coeff} in section A of the Supplemental Material for their form).  We see that the N00N state is consistently the best detector of $SU(2)$ jitter, and that detection probability for any fixed decoherence strength appears to increase (as expected) monotonically with $x$. 

Metrology is also concerned with \emph{estimating} (rather than just detecting) parameters of a process.  In this case, that means estimating $\gamma$, and this requires repeating the experiment more than once, since a single experiment can at best detect that $\gamma>0$.  We can then estimate the probability plotted in Fig. \ref{fig:3}a, e.g. as $\hat{P} = n/N$, where the experiment was repeated $N$ times and decoherence was detected in $n$ of them.  Armed with our knowledge of the initial state and our estimate $\hat{P}$ of the nondetection probability, we can then estimate $\gamma$ (e.g., by simply inverting the appropriate theoretical curve shown in Fig. \ref{fig:3}a.  Of course, our estimate ($\hat{\gamma}$) will have some uncertainty:
$$\gamma = \hat\gamma \pm \Delta\gamma.$$
$\Delta \gamma$ is the smallest change in $\gamma$ that can be detected with reasonable probability.  We refer to it as \emph{sensitivity} (although it should be noted that smaller $\Delta\gamma$ implies greater sensitivity!), and it is given by \cite{okamoto_beating_2008}
\begin{equation} \label{eq:sensitivity}
\Delta \gamma = \frac{\Delta P(\gamma)}{\frac{dP(\gamma)}{d\gamma}},
\end{equation}
where $\Delta P(\gamma)$ is the standard deviation of the \emph{estimated} nondetection probability $\hat{P}$.  Since our detection protocol is a Bernoulli (coin-flip) process, $\Delta P(\gamma)=\sqrt{\frac{P(\gamma)(1-P(\gamma))}{N}}$.  Both $P(\gamma)$ and $\frac{dP(\gamma)}{d\gamma}$ depend only on $x$ (a property of the probe state) and can be computed from Eq. \ref{eq:Ptheory}.

In Fig. \ref{fig:3}b, we show the dependence of sensitivity (scaled by $\sqrt{N}$) on $\gamma$ and the probe state.  Since sensitivity is not a directly observable quantity, we compare a pure theory prediction to an empirical fit.  Dashed lines are pure theoretical predictions, in which both $P(\gamma)$ and $dP(\gamma)/d\gamma$ are calculated using Eq. \ref{eq:Ptheory} for \emph{ideal} input states.  The solid lines are empirical fits:  both $P(\gamma)$ and $dP(\gamma)/d\gamma$ are calculated from the smooth empirical fits to data shown as solid lines in Fig. \ref{fig:3}a.

Whereas N00N states are always the best \emph{detectors} of decoherence, we observe that they are only the best at \emph{measuring} $\gamma$ for low values of $\gamma$.  Around $\gamma\approx1$, they become less sensitive than the other states.  This is a direct consequence of their extreme fragility; since almost any amount of decoherence disturbs the N00N state, it does not distinguish well between medium and strong decoherence.  We also observe a significant discrepancy between the ideal sensitivity (dashed lines) and the observed value at small $\gamma$.  This is because our input states are imperfect and not pure -- even at $\gamma=0$, there is a small probability that we will (falsely) detect decoherence!  Thus, $P(\gamma)$ is never 1, and $\Delta P(\gamma)\not\to0$ as $\gamma\to0$.  Still, we find that N00N states are clearly optimal for $\gamma\lesssim0.9$, beating spin-coherent states by a factor of $1.44\pm0.05$, which agrees well with our theoretical prediction of $\sqrt{2}$ (see section A of the Supplemental Material for derivation).

\section{Quantum Process Tomography}

Using the protocol above, we can \emph{detect} decoherence, and we can even \emph{quantify} its strength.  But to do so reliably, we needed to assume that the process is of a specific one-parameter form (pure $SU(2)$ jitter).  For example, a consistent and coherent $SU(2)$ rotation would violate this assumption, and might go entirely undetected or be incorrectly diagnosed as jitter (depending on the axis of the measurement and the nature of the probe state).  Characterizing general decoherence processes, and correctly diagnosing what is happening, requires quantum process tomography (QPT) \cite{mitchell_diagnosis_2003, obrien_quantum_2004}.

\begin{figure}
\includegraphics[scale=0.5]{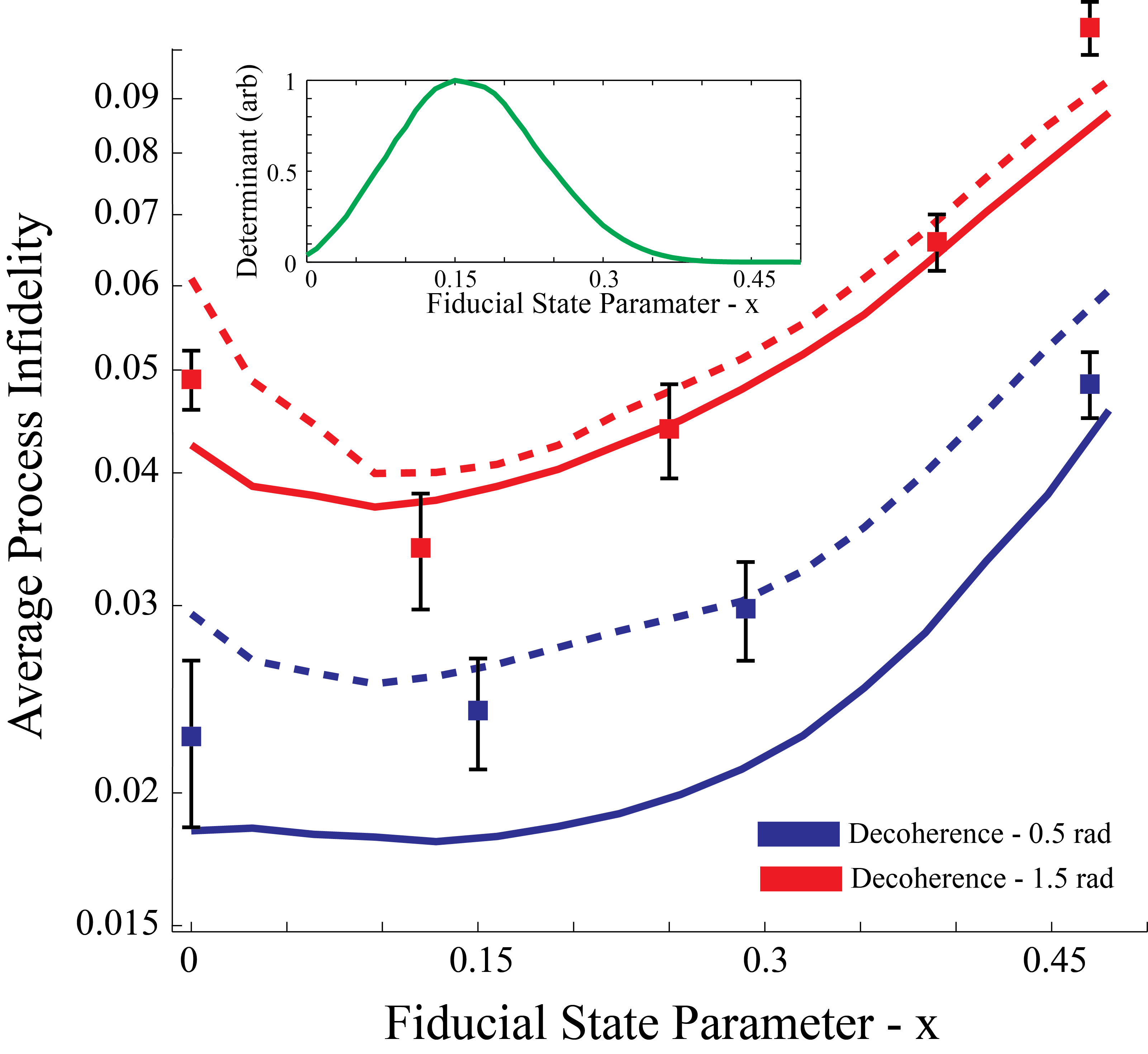}
\caption{\label{fig:4} {\bf Inaccuracy of process reconstruction.}  This figure shows the \emph{average process infidelity} (API, see Eq. \ref{eq:API} -- smaller is better) between the true process and its tomographic reconstruction, and its dependence on the fiducial state used to generate the set of probe states for QPT.  Solid lines are simulations of the experiment using 10 randomly oriented \emph{pure} input states, while dashed lines are simulations using the 10 experimental input states (as determined using state tomography, and in particular accounting for the decline in their purity as $x$ increases).  Squares are experimentally estimated process infidelities.  Red and blue represent different decoherence strengths. {\bf INSET: ``Completeness'' of input sets:}  The inset plot shows the determinant of the probe states' Gram matrix (normalized to a maximum of 1), whose inverse appears in the tomographic reconstruction.  It depends on $x$; larger determinants yield a more robust inversion, while zero determinant indicates a tomographically incomplete set.  We observe that a N00N fiducial state ($x=\frac12$) yields a probe set that is not tomographically complete, while for $x\approx0.15$ the reconstruction should be as robust and accurate as possible.  The main plot confirms this theoretical prediction.}
\end{figure}

QPT requires not one, but an ensemble of input states -- and preparing a suitable ensemble can be quite challenging.  We avoid this complexity by using the ideas of Lobino {\it et al.} \cite{lobino_complete_2008}, and generating diverse input states by applying diverse simple transformations to a single fiducial state (see Fig. \ref{fig:1}d).  Lobino {\it et al.} prepared optical coherent states by translating the $\ket{0}$ state.  The equivalent protocol in our biphoton system would be to prepare the $\ket{2,0}_{H,V}$ state and then generate an ensemble of spin-coherent states by performing different $SU(2)$ rotations on it.  However, we go one step further and generalize this process by varying the fiducial state (parameterized, again, by $x\in[0,1]$).  We study the dependence of process reconstruction fidelity on $x$, to determine (in particular) whether the N00N states that best \emph{detect} decoherence are also the most best probe states to \emph{characterize} it.

QPT reconstructs the entire process matrix (or \emph{superoperator}) from the observed measurement statistics.  This reconstruction, in essence, involves solving a set of linear equations described by a the Gram matrix $M$ of the input states.  The reconstructed process is obtained by applying $M^{-1}$ to a vector of observed statistics.  $M$ must obviously be full rank, but (moreover) it should not have any small eigenvalues.  If $M$ has small eigenvalues, $M^{-1}$ will amplify small statistical fluctuations that result from finite sample size into large errors, and the reconstructed process will have low fidelity with the true process.

A convenient theoretical predictor of a probe ensemble's ``quality'' is the determinant of the corresponding Gram matrix $M$.  Larger determinants are better, implying that the inversion will amplify errors less.  In the inset to Fig. \ref{fig:4}, we plot this determinant as a function of the fiducial state's $x$-value.  For a N00N state ($x=\frac12$) the determinant is zero!  So, remarkably, the set of states generated by the $SU(2)$ orbit of a N00N state is incomplete -- it does not enable QPT at all, and is completely oblivious to at least one parameter of the process (the Gram matrix of the SU(2)-covariant set generated using a N00N state is shown to be rank deficient in section B of the Supplemental Material).  At the opposite extreme ($x=0$), spin-coherent states \emph{do} generate a complete set, suitable for QPT -- but, like the coherent states of Lobino \textit{et al.}, they are not optimal for the task.  The determinant is small, indicating that at least one parameter of the process is poorly resolved.   The maximum value of the determinant is achieved at an intermediate point, $x=\frac{1}{2}-\frac{1}{2\sqrt{2}}\approx0.15$, which generates a set of probe states that are neither N00N nor spin-coherent.   We predicted that this set of states would enable optimally accurate QPT.  

The $x\approx0.15$ ensemble is special and unique in another way; it forms a 2-design (as shown in section B of the Supplemental Material).  Informally, 2-designs are sets of quantum states whose projectors span the vector space of operators as uniformly as possible; more precisely, the ensemble's 2nd moments are equal to those of the uniform Haar ensemble over pure states. Common examples of 2-designs include mutually unbiased bases (MUBs) \cite{wootters_optimal_1989} and symmetric informationally complete measurements (SIC-POVMs) \cite{renes_symmetric_2004,medendorp_experimental_2011}, and there is strong theoretical \cite{wootters_optimal_1989} and experimental \cite{adamson_improving_2010} evidence that 2-designs are optimal for state tomography.  Our results here are the first experimental evidence that 2-designs are optimal for QPT (see theory in Refs. \cite{scott_optimizing_2008,fernandez-perez_quantum_2011}).

Since theoretical analysis predicts that the $x\approx0.15$ ensemble should outperform every other $SU(2)$-generated ensemble at QPT (including the nominally more-sensitive N00N ensemble), we did an experiment to test the prediction.  We prepared several different $SU(2)$-covariant sets of input states -- each generated by applying 10 different collective polarization ($SU(2)$) rotations to a single fiducial state with values of $x$ ranging from $0$ to $0.47$ as detailed in Fig. \ref{fig:1}d -- and used them to perform QPT on an $SU(2)$-jitter process.  \textbf{No a priori assumptions were made about the nature or structure of the process}\footnote{We also verified experimentally that our process preserves the biphoton structure -- i.e., it does not violate permutation symmetry, and therefore does not  change the population in the anti-symmetric subspace -- and so we can treat it as a quantum process on a 3-dimensional Hilbert space.}.  We performed QPT by (1) preparing many copies of each of the 10 states, (2) sending them through the process, (3) performing a tomographically complete set of measurements on each output state, and (4) using maximum-likelihood estimation (MLE) to reconstruct the process.

Evaluating the performance of an experimental tomographic procedure (i.e., to rank our three different input ensembles) is nontrivial.  We cannot assume that we know the ``true'' process, yet the standard metric of tomographic success is ``How close is the reconstructed process to the `true' process?''  We circumvent this problem by using a measure of accuracy that can be estimated directly, the \emph{average process infidelity}(API) \cite{gilchrist_distance_2005}.  The API between two processes $\mathcal{E}$ and $\mathcal{F}$ is the [mixed-state] quantum infidelity $1-F\left(\mathcal{E}\left[\proj{\psi}\right],\mathcal{F}\left[\proj{\psi}\right]\right)$, averaged over all \emph{pure} inputs to the process according to the unitarily invariant Haar measure:
\begin{equation} \label{eq:API}
\mathrm{API}(\mathcal{E},\mathcal{F}) = \int_{\mathrm{Haar}}{\left[1-F\left(\mathcal{E}\left[\proj{\psi}\right],\mathcal{F}\left[\proj{\psi}\right]\right)\right]d\psi}
\end{equation}
The API vanishes for a perfect reconstruction, and increases with errors in tomography.

Once we have used QPT to obtain an estimate $\hat{D}_\gamma$ of the decoherence process, we estimate the API empirically by:
\begin{enumerate}
\item Preparing [many copies of] 40 different randomly chosen input states $\rho_i$ ($i=1\ldots40$).
\item Using state tomography to obtain an estimate $\hat{\rho}_i$ of each input state
\item Applying the decoherence process ($D_\gamma$) to each state.
\item Using state tomography to obtain an estimate $\hat{\rho'}_i$ of each \emph{output} state.
\item Computing the quantum fidelity between (a) the empirical output state $\hat{\rho'}_i$ and (b) the output state \emph{predicted} by our QPT estimate, $\hat{D}_\gamma\left[\rho_i\right]$.
\item Averaging this fidelity over all 40 input states.
\end{enumerate}
The resulting number requires no a priori assertion about the ``true'' process, and it is a good quantifier of how accurately the QPT estimate $\hat{D}_\gamma$ predicts independent experimental results.  But it is also an estimate of the theoretical API as defined in \cite{gilchrist_distance_2005}, and deviates from it only inasmuch as (i) we have approximated the integral in Eq. \ref{eq:API} by a sum over 40 random states; (ii) those states are not quite pure; and (iii) state tomography on finitely many samples is never quite perfect ($\hat{\rho}\neq\rho$).

Figure \ref{fig:4} shows the dependence of the empirical API on the fiducial state parameter $x$, for two different decoherence strengths: $\gamma=0.5$ rad (blue) and $\gamma=1.5$ rad (red).  Points represent experimentally measured APIs, while solid and dashed lines represent two different simulations of our experiment. The solid lines are generated by simulating process tomography using 10 randomly chosen \emph{pure} input states; the dashed lines are generated by simulating process tomography using the same 10 nearly-pure input states used in the experiment.  Both simulations used 40 different random states to estimate the empirical API, just as in the experiment. 

We observe a minimum in the API (i.e., optimal reconstruction fidelity) at $x\approx0.15$ for both of the decoherence strengths -- exactly where theory predicted.  This minimum API coincides with the maximum value of $\mathrm{det}(M)$ (Fig. \ref{fig:4}, inset).  We also confirm that spin-coherent probe ensembles are not optimal.  N00N state ensembles consistently generate the \emph{least} accurate QPT.  Ironically, while the N00N ensemble should in theory fail catastrophically, it is (slightly) redeemed by experimental imperfections in state preparation, which result in the N00N ensemble being not quite perfectly incomplete.  However, it still achieves a much worse API than any other $SU(2)$-covariant input ensemble.

\section{Conclusions} 

It is well known that entanglement can (and usually does) improve metrology.  However, it has also been taken for granted that ensembles of coherent states (which, in multi-photon systems, are not entangled at all) are ``good enough'' for process tomography.  We have shown that both of these beliefs should be interpreted cautiously.  On one hand, while maximally entangled N00N states are indeed optimal for \emph{detecting} a particularly common and important form of decoherence, they are very bad for \emph{characterizing} it in detail.  And while [spin]-coherent states are indeed sufficient for QPT, they are not optimal.  Our results can be summarized as showing that the most robust and flexible way to probe decoherence is with ``partly entangled'' states, intermediate between N00N and coherent states.

Our experimental results show that in the presence of prior information (that the system is undergoing pure $SU(2)$ jitter), the optimal biphoton probe states are N00N states.  We expect that this result will be of utility in magnetometry and atomic physics, where interferometry is often used to estimate noise.  On the other hand, in the complete absence of prior information, we have shown that an intermediate entangled state is much better at performing QPT.  Our method generalizes the technique of Ref. \cite{lobino_complete_2008} -- preparing a single fiducial state and displacing it -- to generate a set of states we believe are optimal for performing QPT on any process.  This set of states forms a 2-design, and our work is the first experimental evidence confirming that 2-designs are optimal for QPT.  Our results imply that one can greatly improve the accuracy of QPT by choosing the right set of input states -- but, surprisingly, the ``right'' states for QPT are not those most sensitive to decoherence.  We conclude that detailed state engineering can be very useful in tailoring probe states or ensembles to specific tasks in the characterization (and ultimately remediation) of decoherence.

We thank NSERC and CIFAR for financial support. We would also like to thank Peter Turner and Steven Flammia for helpful discussions.  Sandia National Laboratories is a multi-program laboratory operated by Sandia Corporation, a wholly owned subsidiary of Lockheed Martin Corporation, for the U.S. Department of Energy's National Nuclear Security Administration under contract DE-AC04-94AL85000.

\bibliographystyle{unsrt}
\bibliography{detecting_a_quantum_process}

\newpage
\section{Supplemental Material}
\subsection{Sensitivity to Decoherence}

The decohering process that we study is described by equation \ref{eq:D}. Then our measurements, projecting the decohered state onto the initial state, can be described as:
\begin{equation}
P(\gamma)=Tr(\ket{\psi_x}\bra{\psi_x}D_\gamma[\ket{\psi_x}\bra{\psi_x}])
\end{equation}
By evaluating the integrals in equation \ref{eq:D} the form presented in equation \ref{eq:pGamma} can be derived.
where $A_x$, $B_x$ and $C_x$ are set by the input state via:
\begin{eqnarray} \label{eq:coeff}
A_x&=&\frac{4}{5}(-x^2+x+\frac{1}{12})\\\nonumber
B_x&=&\frac{8}{15}(x^2-x+\frac{3}{4})\\
C_x&=&\frac{4}{15}(x^2-x+2),\nonumber
\end{eqnarray}
where $x$ is again the ratio of down-converted light to laser light. From these results we can calculate the sensitivity, defined in equation 3, as a function of $x$. Since $P(\gamma)$ is a two outcome projective measurements the uncertainty, due to projection noise, is given by $\Delta P(\gamma) = \sqrt{P(\gamma)(1-P(\gamma))}$ and equation 3, for the sensitivity to decoherence, becomes:
\begin{equation}
\Delta \gamma = \frac{\sqrt{P(\gamma)(1-P(\gamma))}}{\frac{dP(\gamma)}{d\gamma}}.
\end{equation}
Now we can plug in our expressions for $A_x$, $B_x$, $C_x$ (equation 5) and $P(\gamma)$ (equation 2) into equation 6. The result is rather unwieldy, so it is not shown here. But if we evaluate this expression at various values of $\gamma$ we can find the states most sensitive to decoherence. For example, the result of evaluating this at $\gamma=0$ is plotted versus $x$ in figure 5.

From figure 5 it can be seen that there a minimum at $x=0.5$ (N00N state) and a maximum at $x=0$ (spin-coherent state).  Recall that a small $\Delta \gamma$ means that the measurement is more sensitive.  This means that N00N states are the most sensitive 2-photon state to this process. The ratio of the sensitivity of the N00N state to the spin-coherent state is $\sqrt{2}$.
\begin{figure}
\includegraphics[scale=0.5]{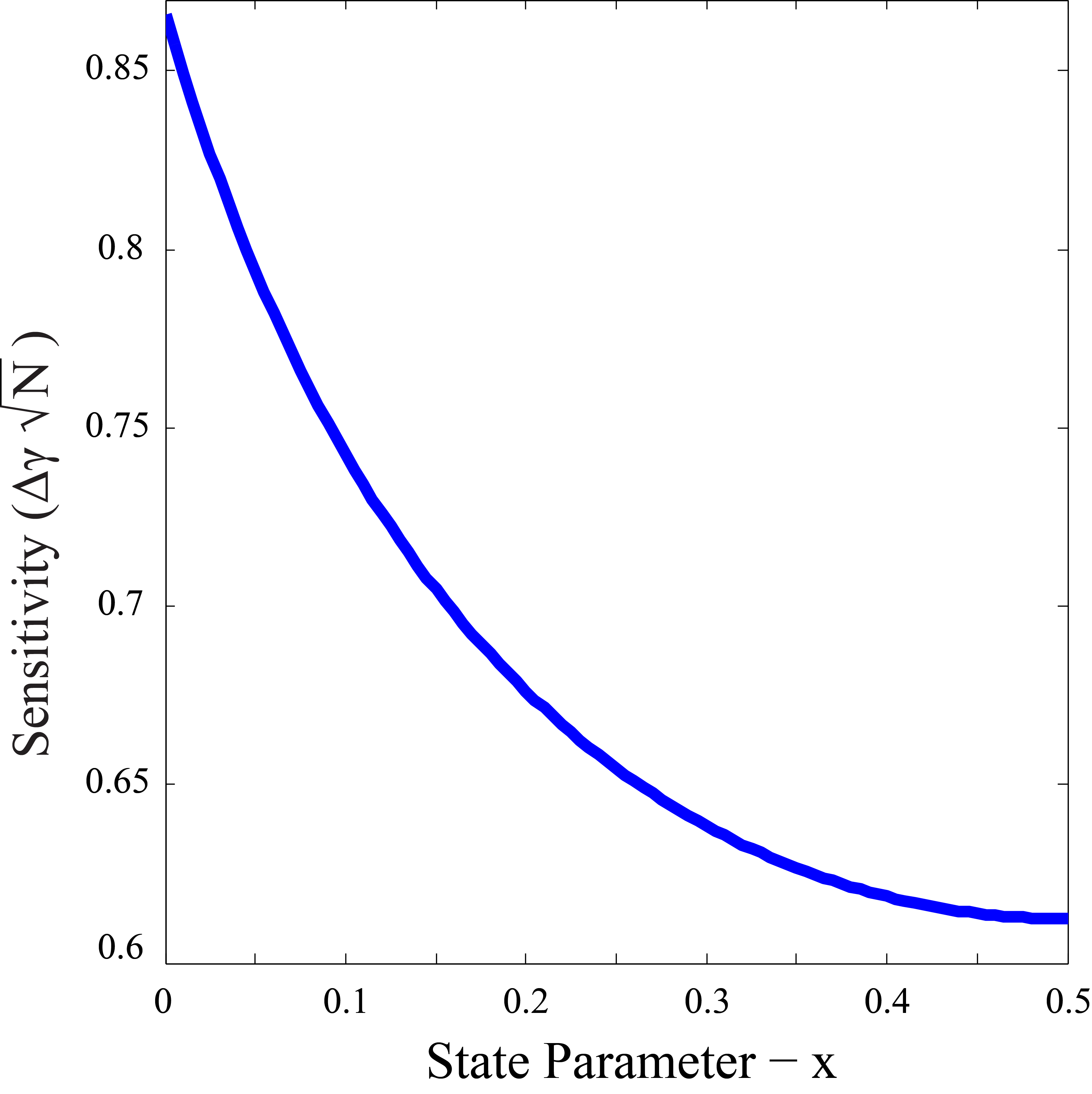}
\caption{\label{fig:5} {\bf Sensitivity of states to small amounts of decoherence}}
\end{figure}

\subsection{Uniformity of $SU(2)$-covariant ensembles}

By an \emph{ensemble}, we mean a set of pure quantum states, equipped with a probability measure.  Some ensembles are discrete, and can be denoted $\{\psi_k:\ k=1\ldots K\}$, where (unless otherwise stated), the associated measure is the \emph{counting measure} that assigns each state probability $1/K$.  But in this paper we are concerned with continuous ensembles, which are uncountable.  We represent such an ensemble by its associated [probability] measure $\mu$, or (to make explicit that the measure is over pure states), $\mu(\psi)$.  Integrals over an ensemble are written as $\int{f(\psi)\mathrm{d}\!\mu(\psi)}.$

At the core of quantum process tomography is inverting Born's rule, which relates observable probabilities to the underlying process:
\begin{equation}
Pr\left( \proj{\psi_{\mathrm{out}}}\right) = \Tr\left( \proj{\psi_{\mathrm{out}}}\mathcal{E}[\proj{\psi_{\mathrm{in}}}] \right).
\end{equation}
This fundamental equation is more elegantly written in Hilbert-Schmidt space (where $\sbraket{A}{B} \equiv \Tr A^\dagger B$ for operators $A$,$B$), as:
\begin{equation}
Pr\left( \proj{\psi_{\mathrm{out}}}\right) =
\sbraopket{\proj{\psi_{\mathrm{out}}}}{\mathcal{E}}{\proj{\psi_{\mathrm{in}}}}.
\end{equation}
These probabilities will identify $\mathcal{E}$ uniquely \emph{only} if both the input and output ensembles independently span the vector space of Hermitian matrices (denoted $\mathcal{B}(\mathcal{H})$ for a Hilbert space $\mathcal{H}$).  In this paper, we use an $SU(2)$-covariant ensemble of input states to probe a process.  If they do not span $\mathcal{B}(\mathcal{H})$ then process tomography will fail outright -- but we also want them to span it \emph{uniformly}.  If every state in the ensemble is \emph{almost} orthogonal to some element $Q$ of $\mathcal{B}(\mathcal{H})$, then we will learn very little about $\mathcal{E}(Q)$, and tomography will be inaccurate.

We can quantify an ensemble's uniformity (at least for this specific purpose) by its ($d^2\times d^2$) Gram matrix,
\begin{equation}
M = \int{\sproj{\proj{\psi}}\mathrm{d}\!\mu(\psi)}. \label{eq:DefGram}
\end{equation}
The Gram matrix has a simple operational meaning; inverting Born's rule involves inverting $M$.  So if $M$ is rank-deficient then inversion is impossible, and if it has small eigenvalues, then finite-sample fluctuations in the estimated probabilities will be amplified by $M^{-1}$ into large errors in the estimated process.  Thus, the most uniform ensemble is the one with the largest minimum eigenvalue.

Since $\Tr M=1$ (by Eq. \ref{eq:DefGram}), the most uniform conceivable ensemble would have $M = \Id/d^2$.  But this is not attainable.  Every quantum state, if viewed as a vector $\sket{\proj{\psi}}$ in matrix space, has a large component of the [Hilbert-Schmidt normalized] identity matrix $\sket{1} = \sket{\Id/\sqrt{d}}$.  Thus
\begin{equation}
\sbraopket{1}{M}{1} = \int{\sbraket{1}{\proj{\psi}} \sbraket{\proj{\psi}}{1}\mathrm{d}\!\psi} = 1/d.
\end{equation}
Subject to this constraint, the most uniform possible ensemble has a Gram matrix given by 
\begin{equation}\label{eq:M2d}
    M = \frac{1}{d}
    \left(\begin{array}{cccc}
				1      & 0             & 0             & \dots \\
			  0      & \frac{1}{d+1} & 0             &  \\
				0      & 0             & \frac{1}{d+1} &  \\
				\vdots &               &               &\ddots 
     \end{array}\right).
\end{equation}
\emph{This} is in fact achievable.  It is achieved by the Haar-uniform ensemble over all pure states, and any other ensemble that achieves it is called a \emph{2-design}.  Such ensembles provide uniform information about every direction in operator space, and their tomographic inversion amplifies experimental noise less than for any non-2-design.

The most elegant way to compute Gram matrices for $SU(2)$-covariant ensembles is using representation theory and Clebsch-Gordan coefficients.  Here, we take a brute-force approach to avoid the mathematical background required for representation theory.

The Gram matrix for an ensemble is given by Eq. \ref{eq:DefGram}.  For our SU(2)-covariant sets generated from spin-1 fiducial states $\ket{\psi_x}$ (equation \ref{eq:psi}) this becomes:
\begin{equation}\label{eq:mx}
M_x=\int{(\hat{U}\otimes\hat{U}^*) (\sproj{\proj{\psi_x}}) (\hat{U}^\dagger\otimes\hat{U}^T)\mathrm{d}\!u},
\end{equation}
where $du$ is the Haar measure, and $u\to\hat{U}$ is the spin-1 representation of SU(2).  We computed Gram matrices explicitly using the Euler representation, where $\hat{U}(\alpha,\beta,\gamma)=e^{-i\alpha\frac{\hat{J}_z}{\hbar}}e^{-i\beta\frac{\hat{J}_y}{\hbar}}e^{-i\gamma\frac{\hat{J}_z}{\hbar}}$ and $du=\frac{3\sin\beta}{8\pi^2}d\alpha~d\beta~d\gamma$.
Although it is involved, this integral can be evaluated for any fiducial state.  We evaluated $M_x$ for three fiducial states and show the resulting (diagonalized) Gram matrices here.

\noindent\textbf{Case 1:} When $x=\frac{1}{2}-\frac{1}{2\sqrt{2}}$, the SU(2) orbit of $\ket{\psi_x}$ forms a 2-design; its Gram matrix is identical to Equation \ref{eq:M2d} for $d=3$:
\begin{equation}
    M_{x=\frac{1}{2}-\frac{1}{2\sqrt{2}}} = \frac{1}{3}\left(\begin{array}{ccccccccc}
				1 & 0 & 0 & 0 & 0 & 0 & 0 & 0 & 0 \\
			  0 & \frac{1}{4} & 0 & 0 & 0 & 0 & 0 & 0 & 0 \\
				0 & 0 & \frac{1}{4} & 0 & 0 & 0 & 0 & 0 & 0 \\
				0 & 0 & 0 & \frac{1}{4} & 0 & 0 & 0 & 0 & 0 \\
				0 & 0 & 0 & 0 & \frac{1}{4} & 0 & 0 & 0 & 0 \\
				0 & 0 & 0 & 0 & 0 & \frac{1}{4} & 0 & 0 & 0 \\
				0 & 0 & 0 & 0 & 0 & 0 & \frac{1}{4} & 0 & 0 \\
				0 & 0 & 0 & 0 & 0 & 0 & 0 & \frac{1}{4} & 0 \\
				0 & 0 & 0 & 0 & 0 & 0 & 0 & 0 & \frac{1}{4} \\
     \end{array}\right).
\end{equation}

\noindent\textbf{Case 2:}  Setting $x=\frac{1}{2}$ makes the fiducial state a N00N state.  For $x=\frac12$, we find that the Gram matrix is rank-deficient:
\begin{equation}
    M_{x=\frac{1}{2}} = \frac{1}{3}\left(\begin{array}{ccccccccc}
				1 & 0 & 0 & 0 & 0 & 0 & 0 & 0 & 0 \\
			  0 & \frac{2}{5} & 0 & 0 & 0 & 0 & 0 & 0 & 0 \\
				0 & 0 & \frac{2}{5} & 0 & 0 & 0 & 0 & 0 & 0 \\
				0 & 0 & 0 & \frac{2}{5} & 0 & 0 & 0 & 0 & 0 \\
				0 & 0 & 0 & 0 & \frac{2}{5} & 0 & 0 & 0 & 0 \\
				0 & 0 & 0 & 0 & 0 & \frac{2}{5} & 0 & 0 & 0 \\
				0 & 0 & 0 & 0 & 0 & 0 & 0 & 0 & 0 \\
				0 & 0 & 0 & 0 & 0 & 0 & 0 & 0 & 0 \\
				0 & 0 & 0 & 0 & 0 & 0 & 0 & 0 & 0 \\
     \end{array}\right),
\end{equation}
confirming that this ensemble is not even tomographically complete.

\noindent\textbf{Case 3:}  Setting $x=0$ makes the fiducial state a spin-coherent state.  This yields a full-rank Gram matrix:
\begin{equation}
    M_{x=0} = \frac{1}{3}\left(\begin{array}{ccccccccc}
				1 & 0 & 0 & 0 & 0 & 0 & 0 & 0 & 0 \\
			  0 & \frac{1}{2} & 0 & 0 & 0 & 0 & 0 & 0 & 0 \\
				0 & 0 & \frac{1}{2} & 0 & 0 & 0 & 0 & 0 & 0 \\
				0 & 0 & 0 & \frac{1}{2} & 0 & 0 & 0 & 0 & 0 \\
				0 & 0 & 0 & 0 & \frac{1}{10} & 0 & 0 & 0 & 0 \\
				0 & 0 & 0 & 0 & 0 & \frac{1}{10} & 0 & 0 & 0 \\
				0 & 0 & 0 & 0 & 0 & 0 & \frac{1}{10} & 0 & 0 \\
				0 & 0 & 0 & 0 & 0 & 0 & 0 & \frac{1}{10} & 0 \\
				0 & 0 & 0 & 0 & 0 & 0 & 0 & 0 & \frac{1}{10} \\
     \end{array}\right),
\end{equation}
but its smallest eigenvalues are significantly smaller than those of $M_{x=\frac{1}{2}-\frac{1}{2\sqrt{2}}}$.  Thus, process tomography using spin-coherent states undersamples certain regions of matrix space, providing relatively little information about the process's action on those operators.  The difference between the uniformity of a 2-design and coherent-state ensemble becomes greater as the system dimension grows.

\end{document}